\documentclass[twocolumn,prb,showpacs,preprintnumbers,amsmath,amssymb]{revtex4}
\usepackage{graphicx}% Include figure files
\usepackage{dcolumn}% Align table columns on decimal point
\usepackage{bm}% bold math
\usepackage{epsfig}
\begin{document}

%\preprint{APS/123-QED}

\title{An asymmetry model for the highly viscous flow}% Force line breaks with \\

\author{U. Buchenau}
\email{buchenau-juelich@t-online.de}
\affiliation{%
Institut f\"ur Festk\"orperforschung, Forschungszentrum J\"ulich\\
Postfach 1913, D--52425 J\"ulich, Federal Republic of Germany}%
\date{July 25, 2009}% It is always \today, today,

\begin{abstract}
The shear flow is modeled in terms of local structural rearrangements. Most of these rearrangements are strongly asymmetric, because the embedding matrix tends to be elastically adapted to the initial state and to have a marked elastic misfit with regard to the final state. As one approaches the Maxwell time, the asymmetry becomes time-dependent, thus enabling the system to leave the initial state. The model explains the Kohlrausch behavior at the main peak in terms of the interaction between different local structural rearrangements.
\end{abstract}


\maketitle

\section{Introduction}

The highly viscous liquid is a theoretical no-man's-land, with many theories and models \cite{heuerr,dyre,adam,ngai,gotz,shoving,diezemann,wolynes,facil,granato,bouchaud,johnson} competing for the explanation of the flow process. The elastic models \cite{dyre} claim a connection between the fragility (the abnormally strong temperature dependence of the main relaxation peak frequency) and the strong temperature dependence of the infinite frequency shear modulus $G$. But this claim has not yet solidified into a detailed microscopic theory.

Experimentally, there is growing evidence \cite{johari,paluch,bohmer,capaccioli} for an intimate connection between the flow process itself (the primary or $\alpha$-peak of the relaxation spectrum) and the faster relaxations three to six decades away. These faster relaxations appear sometimes as a secondary or Johari-Goldstein peak, sometimes just as an excess wing of the primary peak. The experiments support the idea \cite{ngai} that the $\alpha$-peak is a result of the interaction between the "primitive" relaxations on its fast side. But again, this impression has not yet grown into a detailed microscopic theory.

The present paper tries to do a first step towards a more detailed theoretical understanding. The asymmetry model proposed here is based  on another recent experimental finding \cite{olsen} of a pronounced asymmetry of the secondary relaxation. These seem to be relaxation jumps from a lower energy minimum to a higher one, with an energy difference of about 4 $k_BT_g$ at the glass temperature $T_g$.

The asymmetry model ascribes the asymmetry to a structural rearrangement of an inner core, consisting of $N$ atoms or molecules, with an $N$ of the order of twenty to fifty. The structural rearrangement leads to a change of shape, which in turn leads to an elastic distortion, both of the core and of the surrounding matrix. The concept is a good starting point for a quantitative description of the flow process, not only because it provides a quantitative relation between the local flow event and its coupling constant to an external shear, but also because it allows for a simple treatment of the interaction between different relaxing entities.

The following section II is divided in three parts. In the first, the asymmetry is explained in terms of the elastic distortion accompanying a structural rearrangement. Its average value of $4k_BT$ is derived from the statistics of the structures. The second presents a proof of the 1/3-rule for interacting relaxation centers, adapted to the problem. The third derives the shear response for a given energy landscape. Section III gives a comparison to experiment in two cases. Section IV contains the discussion and the conclusions. 

\section{The model}

\subsection{The basic picture}

%%%%%%%%%%%%%%%%%%%%% begin figure %%%%%%%%%%%%%%%%%%%%%%%%%%%%%%%%%%%%%
\begin{figure}[b]
\hspace{-0cm} \vspace{0cm} \epsfig{file=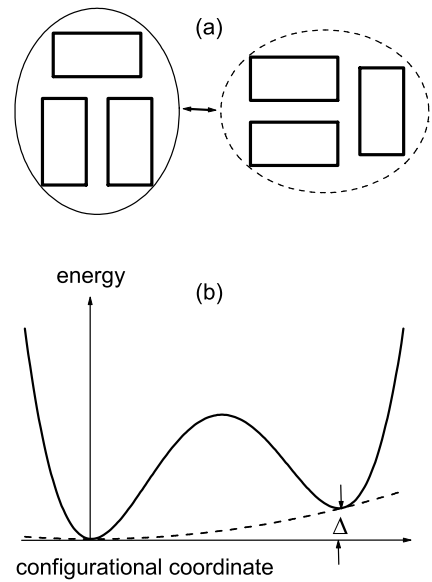,width=6cm,angle=0} \vspace{0cm}\caption{(a) A structural rearrangement of three flat molecules. Note the change of shape. (b) Potential energy, assuming an embedding elastic matrix adapted to the configuration of the left side. The dashed line shows the elastic contribution.}
\end{figure}
%%%%%%%%%%%%%%%%%%%%% end figure %%%%%%%%%%%%%%%%%%%%%%%%%%%%%%%%%%%%%%%

The physical reason for a strong asymmetry of any structural rearrangement is rather obvious. Consider a structural rearrangement of $N$ atoms or molecules with an atomic or molecular volume $v$. Fig. 1 shows as an example the rearrangement of three flat molecules, involving a rotation of one of them by 90 degrees. The figure demonstrates one of the possible relaxation mechanisms for rigid molecules, thus providing a physical explanation for the results of Johari and Goldstein \cite{johari}. Let the surrounding elastic matrix be adapted to the initial state, with no elastic strain of the volume. A transition of the volume $Nv$ to a different stable structure changes the shape of the volume and leads to an elastic misfit, giving rise to an energy difference $\Delta$ between initial and final state. In the following, this mechanism is assumed to dominate the asymmetry of the double-well potentials to such an extent that one can neglect all other influences. The validity of this basic assumption will be discussed in Section IV.

To put the concept on a quantitative basis, assume that the shape difference of the inner volume $Nv$ is either a shear $e_i$ (the shear angle between the two stable structures of the inner volume in radian) or a volume change $\delta v$ or a mixture of both. For a shear $e_i$ and an infinitely hard surrounding medium, the asymmetry (the energy difference between final and initial state) $\Delta$ would be $GNve_i^2/2$, where $G$ is the (infinite frequency) shear modulus. But we consider a surrounding medium with the same elastic constants as the inner part. In this case, the medium distorts as well and one gets a lowering of the elastic energy by about a factor of two, slightly depending on the Poisson ratio of the substance \cite{eshelby}. For our purpose, the difference between the exact factor and two is negligible, so
\begin{equation}\label{d}
    \Delta=\frac{GNve_i^2}{4}
\end{equation}
and the coupling to an external shear $e$ in the direction of $e_i$ is given by
\begin{equation}\label{dde}
    \frac{\partial\Delta}{\partial e}=\frac{GNve_i}{2}.
\end{equation}

One can bring the volume change $N\delta v$ on the same footing as the shear by defining the scaled variable
\begin{equation}
    e_6=\frac{\delta v}{v}\sqrt{\frac{8B}{3B+4G}},
\end{equation}
where $B$ is the infinite frequency bulk modulus and the factor results from the equilibrium between the compressed inner volume $Nv$, which cannot fully expand to its equilibrium value $N(v+\delta v)$, and the expanded surrounding elastic matrix.

With this definition, the asymmetry $\Delta$ resulting from a structural transition of $Nv$ with both shear and volume changes obeys again eq. (\ref{d}), with the square of $e_i$ given by a sum of the squares of its five shear components $e_1$ to $e_5$ and its volume change component $e_6$
\begin{equation}\label{ei}
    e_i^2=\sum_{n=1}^6e_n^2.
\end{equation}
This means a given $\Delta$ defines the surface of a sphere in the six-dimensional distortion space.

Let us next look at the statistics of the possible rearrangements. Taking twenty to fifty atoms or molecules in the inner core, one can estimate the distortion needed for an asymmetry of 4kT. It is relatively small, less than a shear angle of ten degrees. So we are asking for the distribution function of stable structures of the inner core at small distortions. It should be symmetric in the distortion, so it must start with a constant plus a second order term. This second order term will be negative for the volume change, because the average density is preferred. But there is no such preference for a shear.

In the case of non-spherical molecules, each structural rearrangement will be accompanied by a change of molecular orientations, which in turn influence the change of shape. But the distribution function of accessible stable states must be still symmetric in the distortion, starting with a constant at small distortions.

Therefore it is plausible to assume a constant probability density of stable structures in the six coordinates, say $p_{sN}$ in any of the five shear variables and $p_{VN}$ in the volume change $e_6$. The number of states between $e_i$ and $e_i+de_i$ is $\pi^3p_{sN}^5p_{vN}e_i^5de_i$, because the surface of the six-dimensional sphere with radius $e_i$ is $\pi^3e_i^5$. Since $d\Delta=GNve_ide_i/2$, one obtains a number of states $p(\Delta)d\Delta$ between $\Delta$ and $\Delta+d\Delta$
\begin{equation}
    p(\Delta)=\frac{2^9\pi^3p_{sN}^5p_{VN}}{(GNv)^3}\Delta^2\equiv \frac{c_N}{N^3}\frac{\Delta^2}{(Gv)^3},
\end{equation}
where $c_N$ is a dimensionless and temperature-independent number. $c_N$ will tend to increase with $N$, because a larger number of atoms or molecules offers more structural rearrangement possibilities \cite{bouchaud}. $N$ should be decidedly larger than six, because a recent boson peak investigation \cite{schober} in terms of the Eshelby picture showed that the barrier in the case of six atoms or spherical molecules is still close to zero, the elastic restoring forces from the outside being of the same order as the ones from the inside.

The partition function
\begin{equation}\label{zdef}
    Z=1+\int_0^\infty\frac{c_N}{(GNv)^3}\Delta^2{\rm e}^\frac{-\Delta}{k_BT}d\Delta
\end{equation}
is easily evaluated. With the definition of the characteristic multi-minimum parameter $f_N$
\begin{equation}\label{fn}
    f_N=\frac{c_N}{N^3}\left(\frac{k_BT}{Gv}\right)^3,
\end{equation}
the partition function reads
\begin{equation}\label{z}
    Z=1+2f_N.
\end{equation}

The coupling $\gamma$ of a given structural rearrangement with asymmetry $\Delta$ to an external shear strain $e$
\begin{equation}\label{ddde}
    \gamma\equiv\frac{\partial\Delta}{\partial e}=\frac{1}{\sqrt{6}}\frac{\partial\Delta}{\partial e_i}=\pm\sqrt{\Delta GNv/6},
\end{equation}
where the factor $1/\sqrt{6}$ stems from the average over the six possible directions of $e_i$ and the plus or minus sign depends on the orientation of $e$ relative to $e_i$, the plus sign occurring as often as the minus sign. Consequently, the term linear in $e$ in the partition function vanishes. The second order term $Z_e$ results from the integral 
\begin{equation}\label{ze2}
    Z_e=\frac{f_Ne^2GNv}{12k_BT}\int_0^\infty p_{\delta G}(\Delta)d\Delta,
\end{equation}
where the function $p_{\delta G}(\Delta)$ is given by
\begin{equation}\label{deltaG}
	p_{\delta G}(\Delta)=\frac{\Delta^3}{(k_BT)^4}{\rm e}^\frac{-\Delta}{k_BT}.
\end{equation}

The $\Delta$-dependence of the contributions to $\delta G$ is proportional to $\Delta^3\exp(-\Delta/k_BT)$, a function with a maximum at $3k_BT$ and an average $\Delta$ of $4k_BT$. This is close to the asymmetry of 3.8 $k_BT$ found in the key experiment of Dyre and Olsen \cite{olsen}, a first important result of the model.

To second order, the partition function with an applied strain $e$ is
\begin{equation}\label{ze}
    Z=1+2f_N+\frac{f_Ne^2GNv}{2k_BT}
\end{equation}

With eq. (\ref{ze}), one can calculate the contribution $\delta G_N$ of the volume $Nv$ to the reduction of the shear modulus, using the free energy $F=-k_BT\ln Z$ and $\delta G_N=(1/Nv)\partial^2F/\partial e^2$
\begin{equation}\label{dgn}
    \frac{\delta G_N}{G}=-\frac{f_N}{1+2f_N}.
\end{equation}

From eq. (\ref{dgn}), it seems impossible to arrive at the breakdown of the shear modulus, $\delta G/G=-1$. However, as shown in the next subsection, the interaction between different relaxation centers increases their effectivity by a factor of three. Thus one needs only $\delta G/G=-1/3$, which is reached for $f_N=1$.

To understand the physical meaning of a characteristic multiminimum parameter 1, take a single asymmetric double well potential with $\Delta=4k_BT$ within the volume $Nv$, with a coupling constant given by eq. (\ref{ddde}). In this situation $f_N\approx-\delta G/G=1/6\cosh^22\approx 0.012$. This is the value where one crosses over from a single asymmetric double-well to a multiminimum situation. Thus the breakdown of the shear modulus requires hundreds of accessible elastically distorted structural minima around the undistorted ground state, a second important result.

\subsection{The 1/3-rule}

The 1/3-rule, first derived for asymmetric double-well potentials \cite{bujpc}, states that an ensemble of relaxation centers interacting via their coupling to the shear does only need an integrated strength of 1/3 to bring the shear modulus down to zero.

Here one needs a proof of the 1/3-rule for the multiminimum case of $n$ structural minima with energies $U_1, ..U_n$.
Let us define
\begin{equation}
u_i=\frac{U_i}{k_BT}\ \ \ \ \ \frac{\partial u_i}{\partial e}=\gamma_i,
\end{equation}
where $e$ is the external shear strain. The partition function
\begin{equation}
Z=\sum_1^n\exp(-u_i)
\end{equation}
has the first derivative
\begin{equation}
Z'=-\sum_1^n\gamma_i\exp(-u_i)
\end{equation}
and the second derivative
\begin{equation}
Z''=\sum_1^n\gamma_i^2\exp(-u_i)
\end{equation}
with respect to $e$.

The second derivative of $F=-k_BT\ln Z$ can be decomposed into pair contributions $F_{ij}$ with $i\neq j$ \begin{equation}
F_{ij}''=-\frac{k_BT}{Z^2}(\gamma_i-\gamma_j)^2\exp(-u_i-u_j).
\end{equation}

Next one considers the probability $p_i$ to find the system in the structural minimum $i$
\begin{equation}
p_i=\frac{\exp(-u_i)}{Z}
\end{equation}
and its change under the applied shear strain $e$
\begin{equation}
p_i'=-\sum_1^n\frac{\gamma_i-\gamma_j}{Z^2}\exp(-u_i-u_j)
\end{equation}

On application of a small shear strain $e$ at the time zero, the energy $U_{ij}$ transported into the heat bath by jumps between minima $i$ and $j$ in the equilibration of the n-minimum system consists of two terms, a term proportional to $e$ and a second order term
\begin{equation}
U_{ij}^{(2)}=\frac{k_BT}{Z^2}(\gamma_i-\gamma_j)^2\exp(-u_i-u_j)e^2.
\end{equation}
The first order term has to disappear in the integral over the whole sample (otherwise it would not be stable). The second order term is twice the free energy term $F_{ij}''e^2/2$. Since this holds for all pairs $ij$, it holds as well for the whole $n$-minimum system: The equilibration jumps  transport twice the free energy change as heat into the heat bath.

The proof of the 1/3-rule proceeds by noting that the equilibrated $n$-minimum system is still a spanned entropic spring, able to deliver again the energy $F''e^2/2$ to its surroundings. The equilibration of the whole liquid
must convert also this stored energy into heat. Thus each relaxation center delivers three times this energy to the heat bath.

Now consider an ensemble of volume elements $Nv$ with $f_N=1$, which according to the preceding subsection brings the shear modulus down to 2/3 of its infinite frequency value, $\delta G/G=-1/3$. Apply again a small strain $e$ at time zero. In the equilibration, the volumes transport two thirds of the total strain energy to the heat bath. The rest is
lost in the full equilibration of the liquid. Since the energy transported into the heat bath cannot exceed the total strain energy, one concludes that the interaction between different relaxation centers increases their effect on the shear modulus by a factor of three.

Note that the derivation of the 1/3-rule did not require the validity of the basic assumption of the preceding subsection.

The remaining task is to calculate the actual shear response for a given substance. This requires the quantitative treatment of two nontrivial issues: (i) the distribution of relaxation times, determined by the different barriers between structural minima and (ii) the time dependence of the asymmetry, reflecting the interaction between different relaxing entities. An approximation is derived in the next subsection.

\subsection{The shear response for a given energy landscape}

Let us consider the influence of the dynamics. In practice, what matters is not the number of {\it existing} structural alternatives, but rather the number of {\it accessible} structural alternatives. This depends on the relaxation times between the strain-free ground state and the surrounding strained excited states. If a given relaxation time, characterized \cite{cook} by an energy barrier $V$,
\begin{equation}\label{tauv}
    \tau_V=\tau_0\exp\left(\frac{V}{k_BT}\right)
\end{equation}
(with $\tau_0=10^{-13}$ s) allows one to see only one excited state, one is again in the situation characterized by a single double-well potential, even though there may be many other excited states with higher barriers around. If the temperature rises, or if one goes to longer relaxation times, they become visible. The condition $f_N=1$ is reached at a critical barrier $V_c$. The corresponding decay time $\tau_c$ is the time at which the ground state transforms into an excited state of one of the surrounding states.

To quantify the relaxation time influence, one defines a barrier density function $f(V)$ in terms of the relaxational shear modulus decrease $-\delta G/G=f(V)dV$ from all relaxations with barriers between $V$ and $V+dV$. $f(V)$ is related to the conventional \cite{ferry,burelax} relaxation spectrum $H(\tau)
$\begin{equation}\label{rheo}
H(\tau_0{\rm e}^{V/k_BT})=H(\tau_V)=Gk_BTf(V).
\end{equation}

The energy landscape of a given substance is characterized by the barrier density $f_0(V)$ of relaxation entities, which reflects the barrier distribution between the ground state and the excited states. More precisely: If $f_{NV}(V)dV$ is the contribution to $f_N$ from excited states between $V$ and $V+dV$, then $f_0(V)=f_{NV}(V)/3$. For a given $f_0(V)$, $V_c$ can be determined from the 1/3-rule 
\begin{equation}\label{1/3}
	\int_0^\infty f_0(V)\exp(-\tau_V/\tau_c)dV=\frac{1}{3}.
\end{equation}

The final barrier density $f(V)$ results from the interaction of the states $f_0(V)$, which leads to a time dependence of the asymmetry $\Delta$. Let us consider the system in one of the excited states. As long as it does not jump back into the ground state, it is in a state of constant strain. If one holds a piece of matter under a constant shear strain, its elastic energy $\Delta$ decays as $G(t)/G$, where $G(t)$ is the time-dependent shear modulus.

The lowering of $\Delta$ increases the strength of a given relaxation, because the Boltzmann-factor $\exp(-\Delta/k_BT)$ increases. In order to calculate the enhancement for $f_0(V)dV$, one has to do the integration of eq. (\ref{ze2}), limiting oneself to barriers between $V$ and $V+dV$. Therefore the enhancement factor is $(\Delta(0)/\Delta(t))^4$, as it would be for the full integral in eq. (\ref{ze2})
\begin{equation}\label{ff0}
	f(V)=\frac{f_0(V)G^4}{G(\tau_V)^4},
\end{equation}
an equation which should at least be true at low $V$. One has 
\begin{equation}
	\frac{G(\tau_V)}{G}\approx 1-\int_0^V f(V')dV'.
\end{equation}

At low $V$, $f(V)\approx f_0(V)$; the enhancement factor increases with the integral of $f(V)$ over $V$, which at the beginning is close to the integral of $f_0(V)$ over $V$. This suggests the following approximation
\begin{equation}\label{afv}
	f(V)=\frac{f_0(V)\exp(-\tau_V/\tau_c)}{\exp(-4x-\alpha x^2)},
\end{equation}
where
\begin{equation}
	x=\int_0^V f_0(V')\exp(-\tau_V'/\tau_c)dV'.
\end{equation}
The term $-4x$ in the exponent of the enhancement factor of eq. (\ref{afv}) comes from the fourth power in eq. (\ref{ff0}), the coefficient $\alpha$ of the second order term is fixed by the normalization condition
\begin{equation}
	\int_0^\infty f(V)dV=1.
\end{equation}
Trying different $f_0(V)$ to fit substances with and without secondary peak, one always finds $\alpha$-values close to 6.

The approximation provides an elegant way of describing the crossover from the low-barrier region, where one sees the true energy landscape, to the interaction-dominated Kohlrausch behavior at the primary peak. The interaction is correctly taken into account to first order.

%%%%%%%%%%%%%%%%%%%%% begin figure %%%%%%%%%%%%%%%%%%%%%%%%%%%%%%%%%%%%%
\begin{figure}[b]
\hspace{-0cm} \vspace{0cm} \epsfig{file=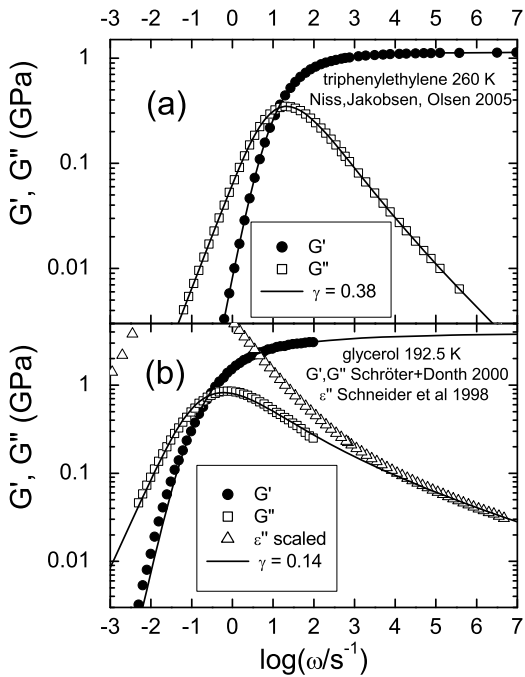,width=7cm,angle=0} \vspace{0cm}\caption{Fit of experimental data (a) in triphenylethylene \cite{niss} at 260 K with $\gamma=0.38$ (b) in glycerol \cite{donth} at 192.5 K with $\gamma=0.14$. In this second case, the fit also reproduces the excess wing of dielectric data \cite{lunkenheimer} at 190 K.}
\end{figure}
%%%%%%%%%%%%%%%%%%%%% end figure %%%%%%%%%%%%%%%%%%%%%%%%%%%%%%%%%%%%%%%

\section{Comparison to experiment}

Eq. (\ref{afv}) allows to extract the energy landscape function $f_0(V)$ from experimental data. This is easiest for substances with an excess wing \cite{roessler,catalin}, which can be fitted with only three parameters: the infinite frequency shear modulus $G$, the decay rate $\tau_c$ and the slope $\gamma$ of the excess wing
\begin{equation}\label{fnv}
    f_0(V)=a\exp\left(\frac{\gamma V}{k_BT}\right),
\end{equation}
where $a$ is the normalization factor required by eq. (\ref{1/3}). Experimentally \cite{catalin}, the exponent $\gamma$ was found to have values around 0.2 in molecular glass formers.

Having $f_0(V)$, one calculates $f(V)$ from eq. (\ref{afv}). $G'(\omega)$ and $G''(\omega)$ are obtained from \cite{burelax}
\begin{equation}
\frac{G'(\omega)}{G}=\int_0^\infty f(V)
\frac{\omega^2\tau_V^2dV}{1+\omega^2\tau_V^2}\label{Gp}
\end{equation}
\begin{equation}
\frac{G''(\omega)}{G}= \int_0^\infty f(V) \frac{\omega\tau_VdV}
{1+\omega^2\tau_V^2}\label{Gpp},
\end{equation}

Fig. 2 (a) shows such a fit for triphenylethylene \cite{niss} at 260 K, 11 K above the glass temperature, with $\tau_c=0.12$ s and $\gamma=0.38$. Fig. 2 (b) shows the best fit to glycerol shear data \cite{donth} about five degrees above the glass transition, with $\tau_c=10.6$ s and $\gamma=0.14$. 

The fit in Fig. 2 (a) is close to the one of a Kohlrausch function with $\beta=0.57$, but its wing has a slightly different slope, namely -0.38 instead of -0.57. This excess is even more pronounced in the glycerol case, where the $\alpha$-peak itself is well fitted by a Kohlrausch law with $\beta=0.43$. Instead, the fit with the asymmetry model provides an excess wing with the much lower slope of 0.14, in good agreement with the one in the scaled dielectric data \cite{lunkenheimer} at 190 K in Fig. 2 (b) (at high frequency, where one sees essentially $f_0(V)$, the dielectric $\epsilon''(\omega)$ should show the same slope as $G''(\omega)$, as long as the coupling constants to dielectrics and shear maintain a constant ratio).

\section{Discussion and conclusions}

Let us first discuss the basic assumption of Section I. It assigns the energy difference between two stable inherent structures to the elastic distortion of the local structure. Obviously, this is not strictly true. One can imagine a structural rearrangement where the elastic distortion is zero. Then, the two energies might still differ, either because the energies of the two core structures differ or because their surface energy to the surrounding matrix differs \cite{wolynes}.

However, neither the structural energy nor the surface energy supplies a quantitative explanation for the high average asymmetry found in experiment. Thus the basic assumption should be essentially correct, at least as an approximation; the other terms are not zero, but their influence on the asymmetry is probably small.

Further support for the basic assumption comes from molecular dynamics results \cite{heuer} which find a pronounced back-jump probability at shorter relaxation times, motivating a description in terms of metabasins, i.e. groups of inherent states which are only left for long relaxation times. Here, this is explained in terms of the elastic distortion, which needs the Maxwell time to decay.

A second question for discussion is the validity of the approximation for the shear response of a given energy landscape in the third part of Section II. It is a crude approximation, in particular for the region of the primary process. Nevertheless, it is a compromise which (i) works and (ii) illustrates an important fact: The shape of the $\alpha$-peak and in particular its slope on the high frequency side are only slightly influenced by the energy landscape and result mainly from the interaction between the relaxing units. One gets a Kohlrausch-$\beta$ of the order of 0.5 even if the underlying energy landscape has a much smaller (or even opposite) slope. Only at much higher frequency one starts again to see the true energy landscape.

In this present version, one has to put in a secondary peak by hand, if there is one in experiment. This is not due to the crude way of treating the effects of the interaction; whenever one has a secondary peak, it remains a peak in the glass phase, where the interaction effects disappear.

To conclude, the asymmetry model presented here describes the flow process in terms of jumps from essentially undistorted ground states into elastically distorted states. The calculated asymmetry agrees with the experimental value. For high enough barriers, the return time becomes so long that the new state has time to relax into a new ground state. At this stage, the undistorted ground state can choose between hundreds of surrounding states. A recipe, based on a rather crude approximation, allows to calculate the shear response for a given energy landscape.

Valuable discussions with Roland B\"ohmer, Kia Ngai, Jeppe Dyre and Andreas Wischnewski are gratefully acknowledged.


\begin{thebibliography}{99}
\bibitem{cook} In principle, the relaxation time for asymmetric conditions is always shorter than the one for the symmetric case given here, also in the multi-minimum case (see G. Williams and M. Cook, Trans. Faraday Soc. {\bf 67}, 990 (1971)), but this is an unnecessary complication for the purpose of this paper.
\end{thebibliography}
\end{document}